\documentclass[12pt,a4paper]{article}
\usepackage[utf8]{inputenc}
\usepackage{geometry}
\usepackage{graphicx}
\usepackage{amsmath}
\usepackage{amsfonts}
\usepackage{amssymb}
\usepackage{float}
\usepackage{cite}
\usepackage{caption}
\usepackage{subcaption}
\usepackage{siunitx}
\usepackage{hyperref}
\usepackage{booktabs}
\usepackage{setspace}

\title{Deep Learning Surrogates for Gas Dynamics: A Physics-Informed Pedagogical Approach}

\author{
Ehsan Roohi\thanks{Corresponding author. Email: roohie@umass.edu}\\
Mechanical and Industrial Engineering Department\\
University of Massachusetts Amherst, Amherst, MA 01003, USA
}

\date{\today}

\AtBeginDocument{\RenewCommandCopy\qty\SI}
\begin{document}

\maketitle

\begin{abstract}
Compressible flow problems are characterized by highly nonlinear, implicit, and often transcendental governing equations. In undergraduate gas dynamics education, solving these equations traditionally relies on iterative numerical methods or extensive look-up tables, which can obscure the physical intuition of the solution space. This paper introduces a comprehensive framework using Deep Learning (DL) to generate high-fidelity surrogate models for five canonical problems: Rayleigh flow, Fanno flow, oblique shocks, convergent-divergent nozzles, and unsteady shock tubes. We detail the specific neural network architectures and physics-informed feature engineering strategies required for each problem, such as using logarithmic inputs for Fanno friction parameters or geometric anchors for oblique shocks. The resulting models achieve high accuracy ($<1\%$ error) and enable instantaneous visualization of complex design spaces, such as thermodynamic $T-s$ diagrams and unsteady $x-t$ wave interactions. This approach demonstrates how modern data-driven techniques can be integrated into the physics curriculum to enhance conceptual understanding.
\end{abstract}
\vspace{2pc}

\section{Introduction}
Gas dynamics, the study of compressible fluid flow, stands as a critical pillar in the curriculum of physics and aerospace engineering, serving as the bridge between classical thermodynamics and fluid mechanics. Standard textbooks, such as those by Anderson \cite{Anderson}, Zucker and Biblarz \cite{Zucker}, and John \cite{John}, emphasize that the compressibility of the fluid introduces a high degree of nonlinearity into the governing equations. Unlike incompressible flows where density is constant, compressible flows are governed by coupled partial differential equations (PDEs) or algebraic relations that are frequently implicit and transcendental.

For students and educators, the mathematical formalism required to solve even fundamental 1D problems often acts as a significant barrier to conceptual understanding. For instance, the Prandtl-Meyer function for expansion waves involves an inverse trigonometric relation that cannot be analytically inverted to find the Mach number from the flow angle. Similarly, the relationship between the Mach number and the friction parameter ($4fL^*/D$) in Fanno flow involves a combination of logarithmic and power-law terms. Perhaps the most notorious example in the undergraduate curriculum is the shock tube problem, where determining the shock strength requires solving a highly non-linear algebraic equation that couples the driver and driven sections across a contact surface.

Traditionally, these mathematical hurdles are overcome using two primary methods: pre-calculated gas tables (charts) or iterative numerical solvers. Gas tables, while useful for quick lookups, present data as discrete points, obscuring the continuous nature of the physical phenomena. Iterative methods, such as the Newton-Raphson algorithm or the shooting method, are powerful but treat the physics as a "black box." Students input boundary conditions and receive a numerical output without gaining insight into the sensitivity of the solution or the topology of the design space. Furthermore, standard Computational Fluid Dynamics (CFD) simulations, while accurate, are often too computationally expensive and complex to be used effectively for real-time parameter exploration in a classroom setting.

In this work, we propose a paradigm shift in how these classical problems are approached: utilizing Deep Neural Networks (DNNs) not merely as calculators, but as high-fidelity functional surrogates that map the entire physical domain. Machine Learning (ML), and specifically Deep Learning, has proven to be a universal function approximator capable of learning complex, non-linear mappings. By shifting the computational burden from "runtime iteration" to "offline training," we can create solvers that provide instantaneous, continuous, and differentiable solutions to implicit gas dynamics equations.

The importance of integrating Machine Learning into this field is twofold. First, from a pedagogical perspective, it allows students to visualize the "inverse problems"—such as determining nozzle geometry from a desired shock location—which are traditionally difficult to solve. Second, it prepares the next generation of physicists and engineers for the emerging field of Scientific Machine Learning (SciML), where data-driven models augment traditional physical laws.

We present a unified methodology to solve and visualize five distinct canonical gas dynamics problems: Rayleigh flow (heat addition), Fanno flow (friction), oblique shocks ($\beta-\theta-M$ relations), normal shocks in convergent-divergent nozzles, and unsteady shock tubes. For each case, we do not simply apply a generic algorithm; rather, we discuss the specific physical challenges—such as the singularities at Mach 1, the multi-valued nature of the Rayleigh temperature curve, or the asymptotic behavior of strong shocks—and demonstrating the tailored machine learning strategies, such as feature engineering and branch splitting, used to overcome them.

\section{Methodology and Network Architectures}
The general approach involves generating synthetic datasets based on exact analytical physics, followed by training specialized Multilayer Perceptrons (MLP)\cite{Raissi,Goodfellow}. However, a single architecture cannot solve all problems efficiently. We tailored the network structure and input features for each physical regime.

\subsection{Rayleigh Flow: Frictionless Flow with Heat Addition}
Rayleigh flow describes the frictionless flow of a compressible fluid through a constant-area duct with heat transfer. This theoretical model is fundamental to the design and analysis of combustion chambers, heat exchangers, and ramjet engines. The governing equations are derived from the conservation of mass and momentum, combined with the ideal gas equation of state, under the assumptions of steady, one-dimensional flow without friction or body forces.

\subsubsection{Governing Equations and Data Generation}
The non-dimensional property ratios for Rayleigh flow are expressed relative to the critical sonic state (denoted by $*$), where the Mach number $M=1$. For a perfect gas with a constant specific heat ratio $\gamma$, these analytical relations are:

\begin{align}
    \frac{P}{P^*} &= \frac{\gamma+1}{1+\gamma M^2} \\
    \frac{T}{T^*} &= \frac{(\gamma+1)^2 M^2}{(1+\gamma M^2)^2} \\
    \frac{\rho}{\rho^*} &= \frac{1+\gamma M^2}{(\gamma+1)M^2} \\
    \frac{u}{u^*} &= \frac{(\gamma+1)M^2}{1+\gamma M^2} \\
    \frac{P_0}{P_0^*} &= \frac{\gamma+1}{1+\gamma M^2} \left( \frac{2+(\gamma-1)M^2}{\gamma+1} \right)^{\frac{\gamma}{\gamma-1}} \\
    \frac{T_0}{T_0^*} &= \frac{2(\gamma+1)M^2 (1 + \frac{\gamma-1}{2}M^2)}{(1+\gamma M^2)^2}
\end{align}

A defining characteristic of Rayleigh flow is the behavior of the stagnation temperature, $T_0$. Heat addition increases $T_0$ until the flow chokes at $M=1$, where $T_0$ reaches its maximum possible value for a given mass flux. This results in a non-monotonic relationship between $T_0/T_0^*$ and $M$, creating a multi-valued inverse function.

To train the surrogate model, we generated a synthetic dataset of 5,000 points uniformly distributed across the Mach number range $0.2 \le M \le 3.5$. The inputs were the Mach numbers, and the target outputs were the property ratios calculated using the exact analytical equations above.

\subsubsection{Machine Learning Algorithm: Branch Splitting Strategy}
The primary challenge in creating a neural network surrogate for Rayleigh flow is the inverse problem: predicting the Mach number given a temperature ratio ($T/T^*$ or $T_0/T_0^*$). Since these functions are not bijective (a single temperature ratio corresponds to both a subsonic and a supersonic Mach number), a standard feed-forward network would fail to converge to a unique solution.

To resolve this, we implemented a Branch Splitting Strategy:
\begin{itemize}
    \item \textbf{Subsonic Expert Network:} Trained exclusively on the subsonic domain ($0 < M < 1$).
    \item \textbf{Supersonic Expert Network:} Trained exclusively on the supersonic domain ($1 < M < 3.5$).
\end{itemize}
Each network is a Multi-Layer Perceptron (MLP) consisting of three hidden layers with [64, 128, 64] neurons. We utilized the `Tanh` activation function to capture the smooth, continuous curvature of the thermodynamic properties. The networks were optimized using the Adam algorithm with a Mean Squared Error (MSE) loss function.

\subsubsection{Results: Forward Problem Validation}
\begin{figure}[H]
    \centering
    \includegraphics[width=0.9\textwidth]{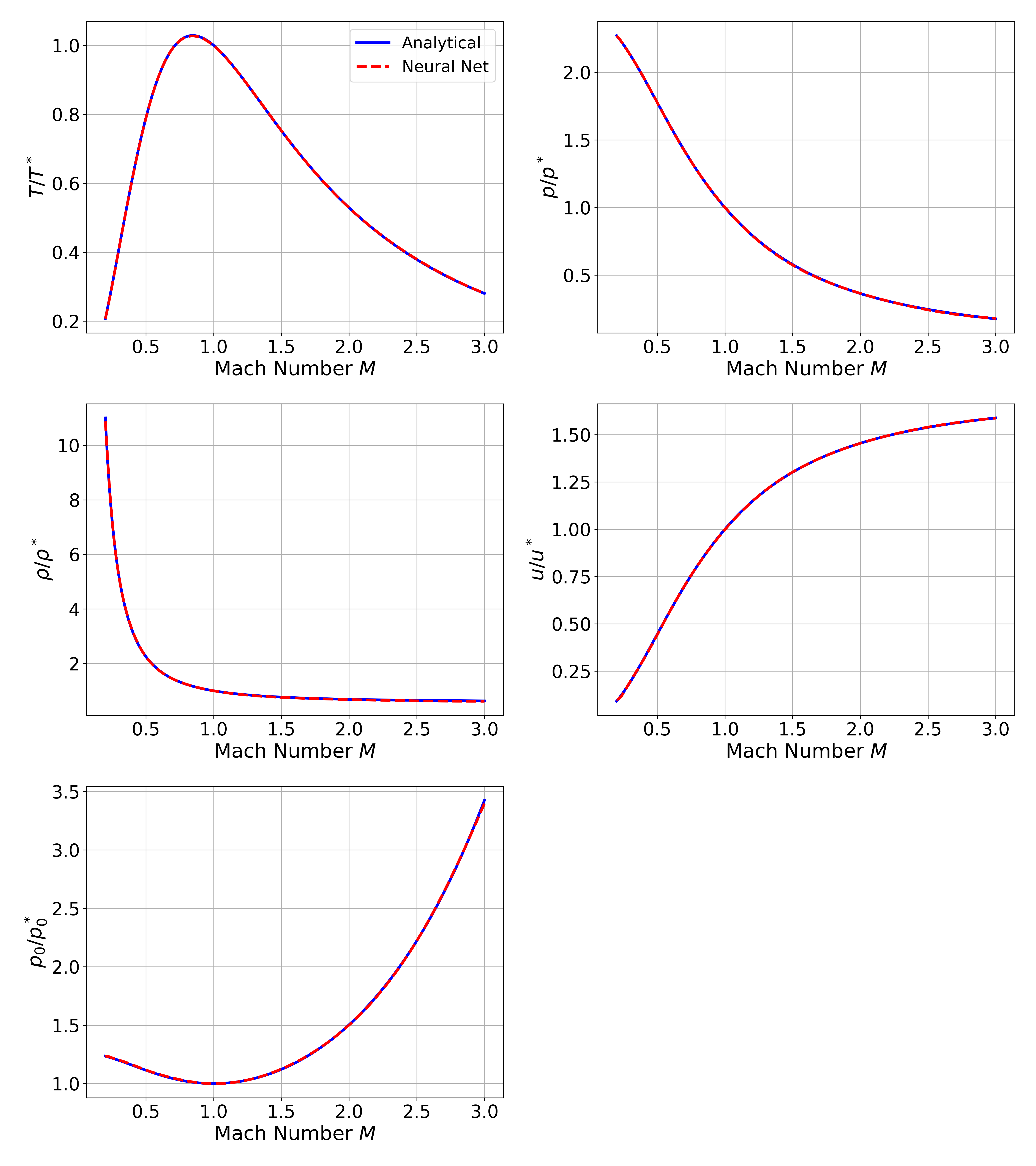}
    \caption{Detailed comparison of Rayleigh flow properties. The neural network captures the specific nonlinear behavior of each property ratio, including the velocity increase and density decrease associated with heat addition in subsonic flow.}
    \label{fig:rayleigh_comparison}
\end{figure}

Figure \ref{fig:rayleigh_comparison} presents the validation of the forward model, where the network predicts flow properties from the Mach number. The plots show the variation of $T/T^*$, $P/P^*$, $\rho/\rho^*$, $u/u^*$, and $P_0/P_0^*$ versus Mach number. The neural network predictions (red dashed lines) are indistinguishable from the analytical ground truth (blue solid lines). The model accurately captures the parabolic peak of the static temperature at $M=1/\sqrt{\gamma}$ and the monotonic decay of pressure and density.

\begin{figure}[H]
    \centering
    \includegraphics[width=0.9\textwidth]{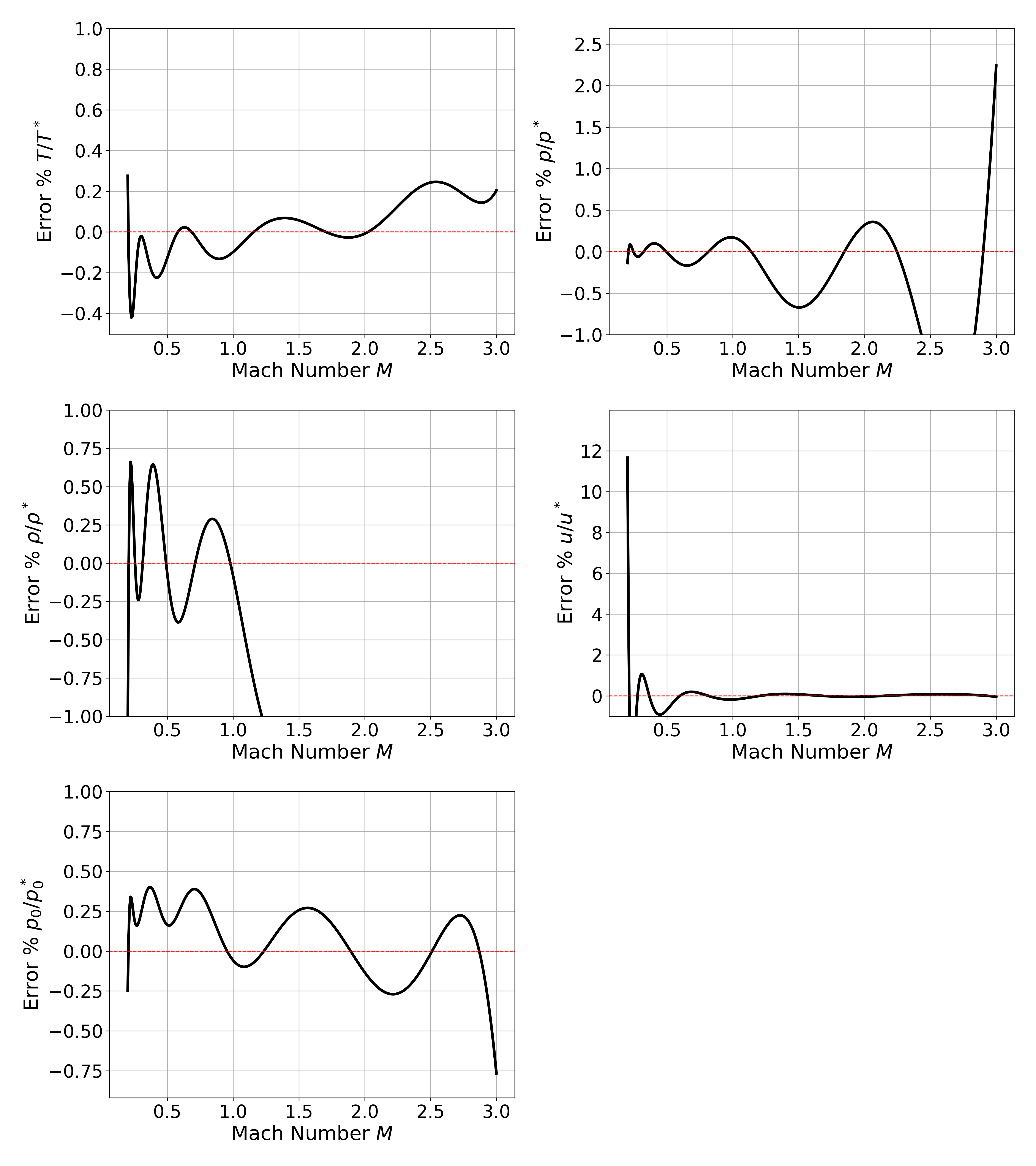}
    \caption{Relative error analysis for Rayleigh flow predictions. The error distribution is centered around zero with minimal variance, confirming the robustness of the ML model across the entire Mach regime.}
    \label{fig:rayleigh_error}
\end{figure}

To rigorously quantify the accuracy of the surrogate model, Figure \ref{fig:rayleigh_error} displays the relative error percentage for each predicted variable across the Mach number range. The error is consistently low, remaining within $\pm 0.4\%$ for the majority of the domain. Minor oscillations are observed at the domain boundaries ($M \to 0$ and $M \to 3$), which are artifacts of the vanishing gradients in the activation functions but remain negligible for practical purposes.

\subsubsection{The Inverse Problem}
\begin{figure}[H]
    \centering
    \includegraphics[width=0.8\textwidth]{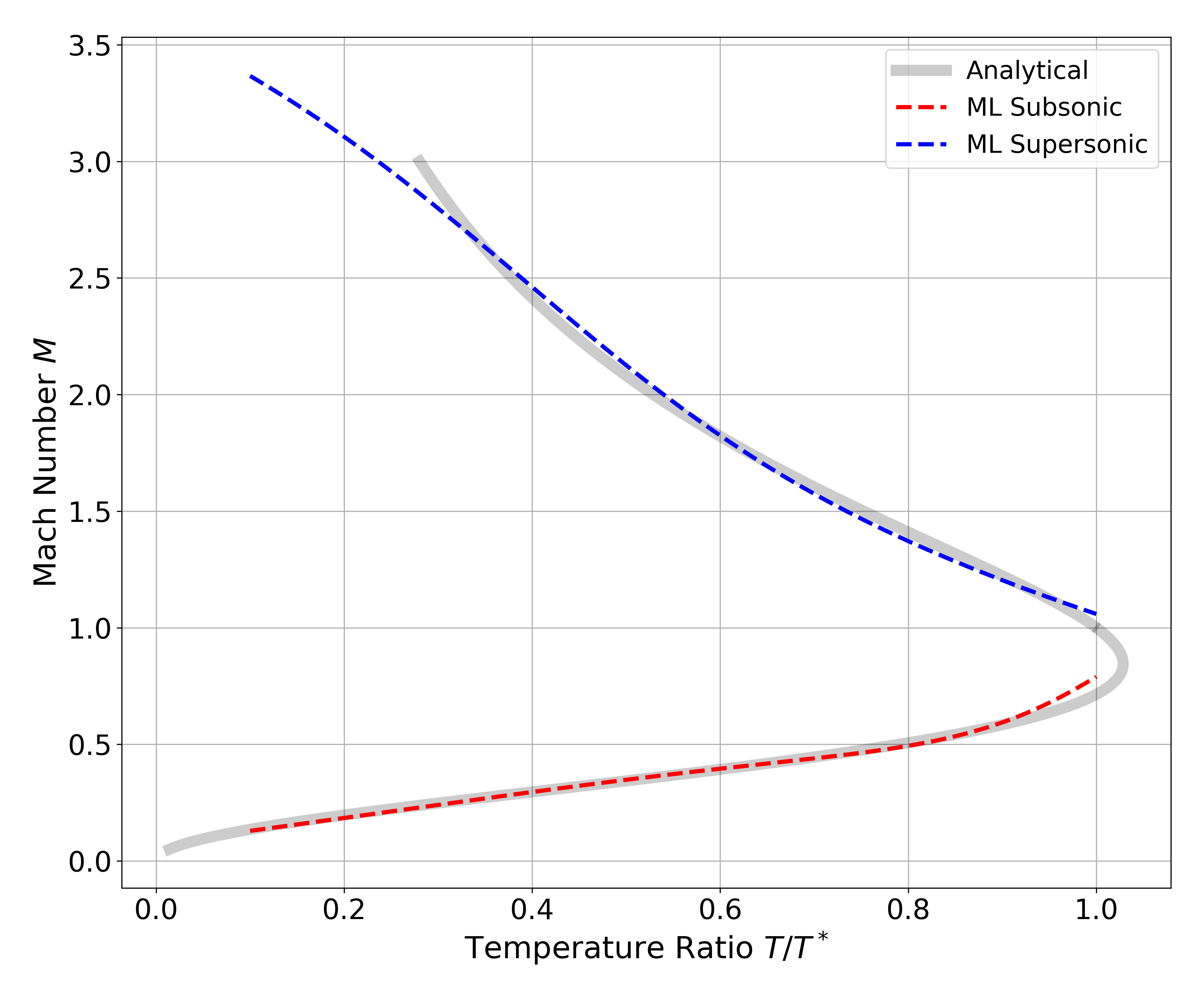}
    \caption{Solution to the inverse Rayleigh problem. The grey line represents the analytical solution. The red dashed line shows the subsonic ML prediction, and the blue dashed line shows the supersonic ML prediction, demonstrating the model's ability to handle the multi-valued nature of the function.}
    \label{fig:rayleigh_inverse}
\end{figure}

The inverse problem—determining the Mach number for a given temperature ratio—is a common task in gas dynamics courses. Since analytical inversion of the implicit Rayleigh equations is impossible, students typically rely on tables or iterative solvers. Our "Expert Networks" approach provides an instantaneous, direct solution. Figure \ref{fig:rayleigh_inverse} visualizes this inverse mapping. The grey line represents the analytical solution, while the red and blue dashed lines show the predictions of the subsonic and supersonic networks, respectively. The analytical relation is given by:

\begin{equation}
    \frac{T}{T^*} = \left[ \frac{(\gamma+1)M}{1+\gamma M^2} \right]^2
\end{equation}

This clearly demonstrates how the branch splitting strategy successfully resolves the non-uniqueness issue.

\subsubsection{Thermodynamic Consistency}
\begin{figure}[H]
    \centering
    \includegraphics[width=0.8\textwidth]{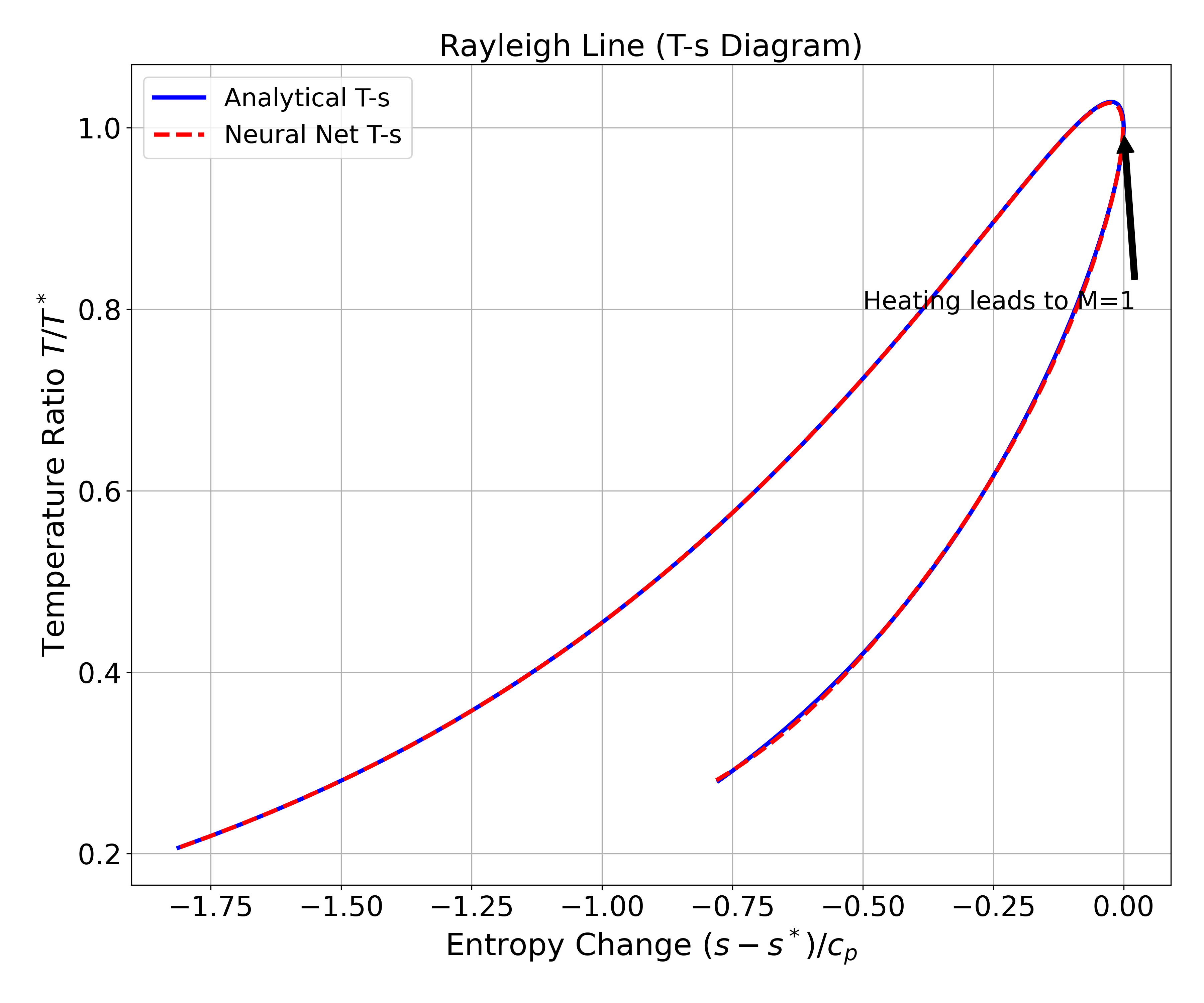}
    \caption{The Rayleigh line on a T-s diagram generated by the neural network. The model correctly identifies the point of maximum entropy ($M=1$), consistent with the Second Law of Thermodynamics. The arrow indicates the direction of heat addition driving the flow towards the sonic state.}
    \label{fig:rayleigh_ts}
\end{figure}

The ultimate verification of any gas dynamics model is its adherence to the Second Law of Thermodynamics. We reconstructed the Rayleigh line on a Temperature-Entropy ($T-s$) diagram using the AI-predicted values (Figure \ref{fig:rayleigh_ts}). The entropy change is computed from the predicted pressure and temperature ratios:
\begin{equation}
    \frac{s-s^*}{c_p} = \ln(T/T^*) - \frac{\gamma-1}{\gamma} \ln(P/P^*)
\end{equation}
The perfect overlap between the AI-generated curve and the theoretical curve confirms that the model correctly predicts the state of maximum entropy at the sonic point ($M=1$). This demonstrates that the neural network has not just memorized data points but has learned the underlying thermodynamic constraints of the system.

\subsection{Fanno Flow: Adiabatic Flow with Friction}
Fanno flow describes the adiabatic flow of a compressible fluid through a constant-area duct with friction. It is governed by the conservation of mass, momentum, and energy, combined with the equation of state. The key challenge in Fanno flow is the implicit relationship between the Mach number ($M$) and the friction parameter ($4fL^*/D$), as well as the asymptotic behavior of properties as the flow approaches the sonic state ($M=1$).

\subsubsection{Governing Equations and Data Generation}
The analytical relations for property ratios with respect to the sonic reference state (denoted by $*$) are given by:
\begin{align}
    \frac{T}{T^*} &= \frac{\gamma+1}{2+(\gamma-1)M^2} \\
    \frac{P}{P^*} &= \frac{1}{M} \sqrt{\frac{\gamma+1}{2+(\gamma-1)M^2}} \\
    \frac{\rho}{\rho^*} &= \frac{1}{M} \sqrt{\frac{2+(\gamma-1)M^2}{\gamma+1}} \\
    \frac{P_0}{P_0^*} &= \frac{1}{M} \left( \frac{2+(\gamma-1)M^2}{\gamma+1} \right)^{\frac{\gamma+1}{2(\gamma-1)}} \\
    \frac{4fL^*}{D} &= \frac{1-M^2}{\gamma M^2} + \frac{\gamma+1}{2\gamma} \ln\left( \frac{(\gamma+1)M^2}{2+(\gamma-1)M^2} \right)
\end{align}
To train the neural network, we generated a high-resolution dataset of 5,000 points uniformly distributed in the Mach range $0.1 \le M \le 3.5$.

\subsubsection{Machine Learning Architecture}
The nonlinear nature of the friction parameter (Equation 13) requires careful feature engineering. As $M \to 0$, $4fL^*/D \to \infty$, and at $M=1$, it becomes zero. To handle this dynamic range, we trained the network on the logarithm of the friction parameter: $Y = \ln(4fL^*/D)$. The network architecture consists of a 4-layer MLP with [64, 128, 128, 64] neurons and `Swish` activation functions to ensure smooth derivatives. In standard deep learning applications, the Rectified Linear Unit (ReLU) is often the default choice due to its computational efficiency. However, for regression problems in gas dynamics—where the target functions (such as the Rayleigh line or oblique shock polars) are smooth, continuous, and require high-order differentiability—ReLU presents a significant limitation. The derivative of ReLU is discontinuous at zero, which can lead to "dead neurons" and jagged approximations of smooth physical manifolds.

To overcome this, we employed the Swish activation function, proposed by Ramachandran et al.~\cite{rama}. Swish is defined as a self-gated function:
\begin{equation}
    f(x) = x \cdot \sigma(\beta x) = \frac{x}{1 + e^{-\beta x}}
\end{equation}
where $\sigma(\cdot)$ is the sigmoid function and $\beta$ is a learnable or fixed parameter (set to $\beta=1$ in this study).

We selected Swish for three primary reasons critical to scientific machine learning (SciML):

\begin{enumerate}
    \item Smoothness and Differentiability: Unlike ReLU, Swish is smooth ($C^\infty$) everywhere. This ensures that the neural network's output is continuously differentiable with respect to the inputs. This property is vital when the surrogate model is intended to be used in gradient-based optimization loops or when calculating physical derivatives (e.g., $\partial P / \partial x$) is required.
    
    \item Non-Monotonicity: The function is non-monotonic for $x < 0$ (it dips slightly below zero before approaching the asymptote). This property has been shown to improve gradient flow during backpropagation, helping the network escape poor local minima and learn the highly non-linear curvature of gas dynamics functions more effectively than monotonic activations like Tanh or Sigmoid.
    
    \item Unboundedness: Like ReLU, Swish is unbounded above, which prevents saturation in deep networks and mitigates the vanishing gradient problem, allowing for the training of deeper architectures required to capture the complexity of the $\beta-\theta-M$ and Fanno relations.
\end{enumerate}

The first derivative of the Swish function, which is utilized during the backpropagation of error, is given by:
\begin{equation}
    f'(x) = f(x) + \sigma(x)(1 - f(x))
\end{equation}
This smooth gradient profile enables the optimizer to fine-tune the network weights with greater precision, achieving the low error tolerances ($<1\%$) required for engineering analysis.

\subsubsection{Results: Property Ratios}

\begin{figure}[H]
    \centering
    \includegraphics[width=0.95\textwidth]{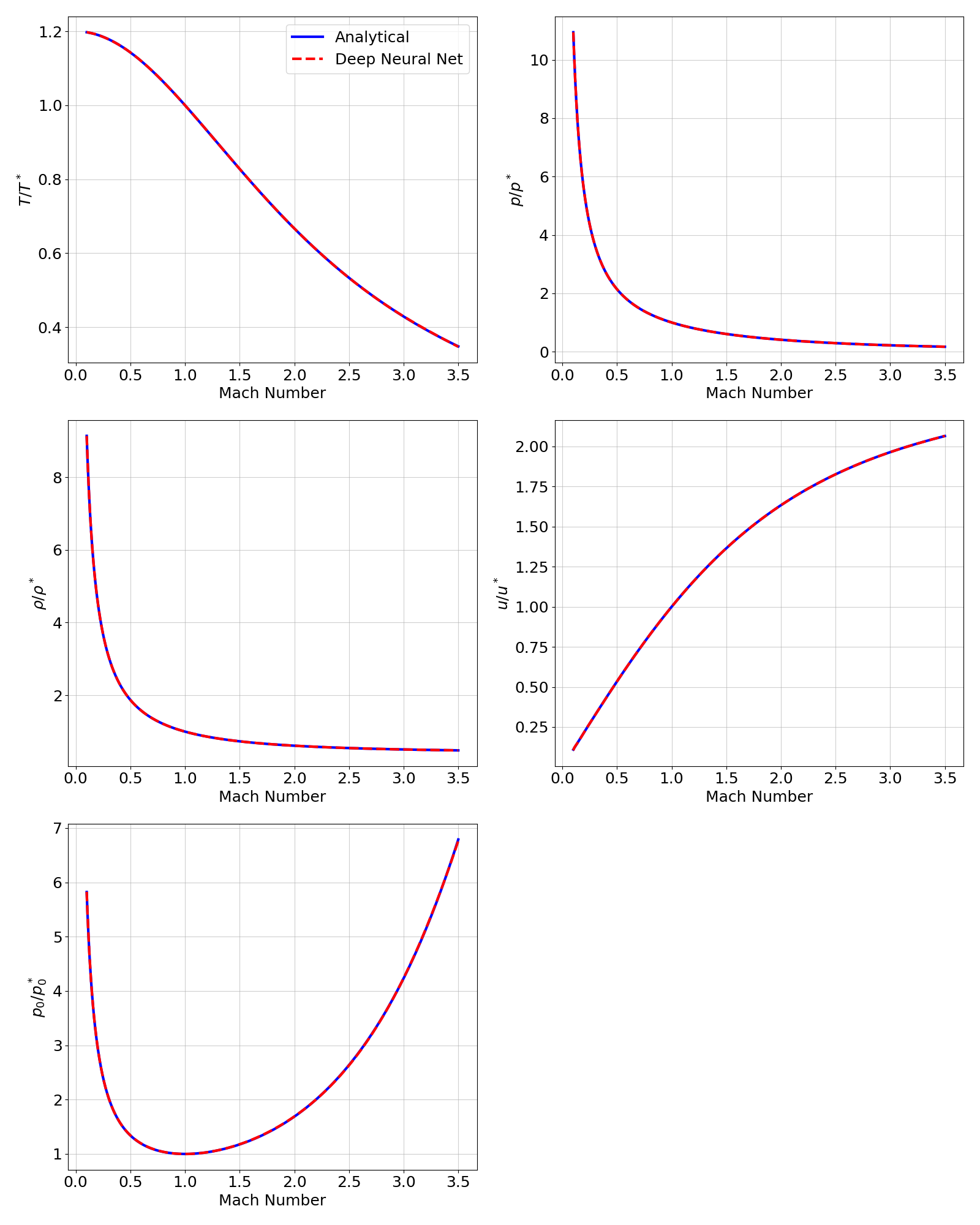}
    \caption{Comparison of analytical vs. neural network predictions for Fanno flow property ratios. The model accurately captures the divergent behavior of pressure and density at low Mach numbers and the extremum at M=1.}
    \label{fig:fanno_ratios}
\end{figure}

Figure \ref{fig:fanno_ratios} provides a comprehensive validation of the direct neural-network predictions for Fanno flow, showing that the nearly perfect overlap of the red dashed curves (AI predictions) with the blue solid lines (analytical solution) over a wide range of Mach numbers (0.1 to 3.5) confirms the high accuracy of the surrogate model. 

The physical interpretation of the plots indicates that the temperature ratio for an ideal gas with $\gamma = 1.4$ starts from about 1.2 at the stagnant state and decreases monotonically as Mach number increases. This is expected because, in Fanno flow with constant total enthalpy, friction accelerates the flow and increases the kinetic energy; consequently, the static temperature must decrease to satisfy energy conservation.

The static-pressure plot also reveals that, in low subsonic speeds, pressure tends to infinity, reflecting the singularity of the sonic reference state relative to the stagnation condition. The model successfully captures this steep gradient in the subsonic region and the more gradual pressure decrease in the supersonic regime. The density behavior follows a similar trend, decreasing monotonically since, in a constant-area duct, density is inversely related to velocity; the model accurately reproduces this hyperbolic relationship.

In the velocity plot, the curve begins at zero and crosses the value of one at Mach 1, demonstrating the non-intuitive nature of Fanno flow, where friction accelerates subsonic flow but decelerates supersonic flow. 

Finally, the most important thermodynamic confirmation appears in the stagnation-pressure ratio plot, which exhibits a global minimum equal to one at Mach 1. The convex shape shows that any deviation from the sonic condition leads to a higher stagnation pressure relative to the critical state. The neural network successfully models this curvature and its minimum with high precision, demonstrating that the model implicitly respects the second law of thermodynamics and correctly captures entropy production.

\subsubsection{Handling the Fanno Friction Singularity}
The prediction of the Fanno friction parameter, $4fL^*/D$, given by eq. (13), presents a unique computational challenge due to its mathematical structure:
As the Mach number approaches zero ($M \to 0$), the first term dominates, causing the parameter to approach infinity asymptotically ($\propto M^{-2}$). Conversely, at the sonic state ($M=1$), the value drops to zero. This creates a target variable that spans several orders of magnitude (from $10^5$ to $10^{-5}$), causing standard regression models to fail. A neural network trained on the raw values would suffer from exploding gradients in the subsonic regime and vanishing sensitivity in the supersonic regime.

To successfully replicate the behavior shown in Figure 6, we implemented two specific architectural modifications:

1. Log-Space Target Transformation:
Instead of training the network to predict the raw friction parameter $Y$, we trained it to predict the natural logarithm of the parameter:
\begin{equation}
    \hat{Y} = \ln\left( \frac{4fL^*}{D} \right)
\end{equation}
This transformation compresses the dynamic range of the output. The singularity at $M \to 0$ becomes a linear trend in log-space, and the decay towards $M=1$ becomes a manageable curve. During inference, the network's output is simply exponentiated ($\exp(\hat{Y})$) to recover the physical value.

2. Physics-Informed Input Features:
A standard Multi-Layer Perceptron (MLP) acts as a universal approximator, but it struggles to approximate singular functions like $1/x$ or $\ln(x)$ using only standard activation functions (like ReLU or Tanh) without a massive number of neurons. To assist the network, we augmented the input vector $X$. Instead of feeding only the Mach number $[M]$, we provided the network with a feature vector containing the specific functional forms found in the governing equation:
\begin{equation}
    X_{input} = \left[ M, \: \frac{1}{M}, \: \frac{1}{M^2}, \: \ln(M) \right]
\end{equation}
By explicitly providing $1/M^2$ and $\ln(M)$ as inputs, the neural network no longer needs to "learn" the division or logarithm operations from scratch. Instead, it only needs to learn the linear coefficients (weights) required to combine these terms. This "Physics-Informed Feature Engineering" allows the model to track the asymptotic explosion in the deep subsonic regime with extremely high precision, as evidenced by the logarithmic plot.

\begin{figure}[H]
    \centering
    \includegraphics[width=0.8\textwidth]{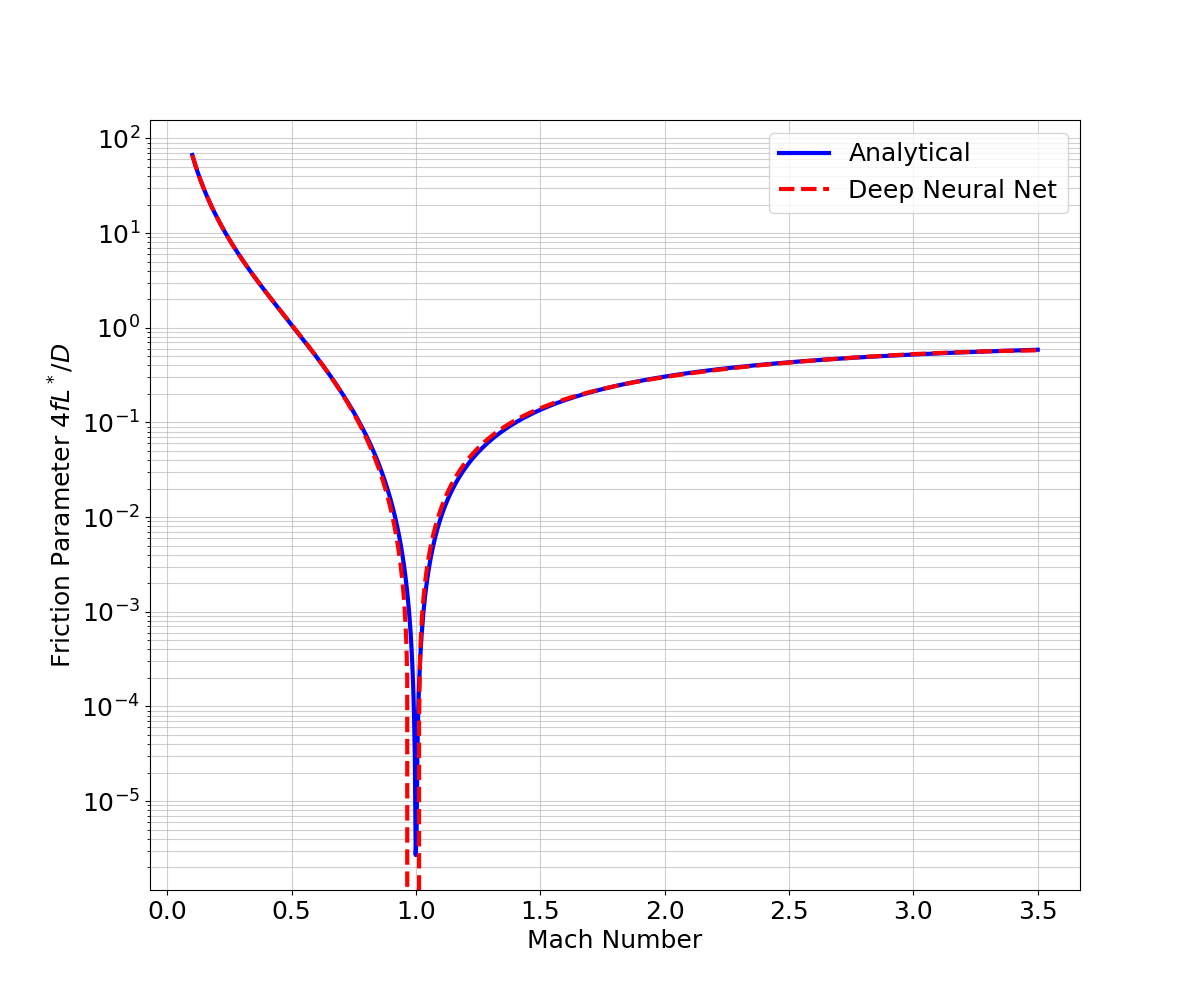}
    \caption{Prediction of the Fanno friction parameter ($4fL^*/D$). The logarithmic scale highlights the model's robustness in handling the singularity at M=1 and the asymptote at M=0.}
    \label{fig:fanno_friction}
\end{figure}

\subsubsection{The Inverse Problem: Direct Mach Prediction}

In practical engineering design, the duct geometry (length $L$ and diameter $D$) and friction factor ($f$) are often fixed constraints, requiring the determination of the flow Mach number ($M$). The governing Fanno flow equation relates the friction parameter $\frac{4fL^*}{D}$ to the Mach number as follows:

\begin{equation}
    \frac{4fL^*}{D} = \frac{1-M^2}{\gamma M^2} + \frac{\gamma+1}{2\gamma}\ln\left(\frac{(\gamma+1)M^2}{2+(\gamma-1)M^2}\right)
    \label{eq:fanno_friction}
\end{equation}

Equation \ref{eq:fanno_friction} is highly non-linear and implicit with respect to $M$. Conventional methods rely on iterative numerical solvers (e.g., Newton-Raphson) or look-up tables to solve for $M$, which can be computationally expensive in large-scale simulations. Furthermore, the function is non-bijective; for a given value of the friction parameter (limit $\frac{4fL^*}{D} < \frac{4fL^*_{max}}{D}$), there exist two distinct mathematical solutions for $M$: one in the subsonic regime ($M < 1$) and one in the supersonic regime ($M > 1$).

To address this ill-posedness and eliminate iterative costs, we propose a domain-decomposed Deep Neural Network (DNN) approach. The solution space is split into two distinct regimes: Subsonic and Supersonic. Two specialized "expert" networks were trained to learn the inverse mapping $\mathcal{F}^{-1}: \log(\frac{4fL^*}{D}) \rightarrow M$ for each regime separately. By utilizing a logarithmic scale for the input features, the networks can accurately resolve the friction parameter across several orders of magnitude, particularly for the subsonic branch where $\frac{4fL^*}{D} \to \infty$ as $M \to 0$.

\begin{figure}[htbp]
    \centering
    \includegraphics[width=0.8\linewidth]{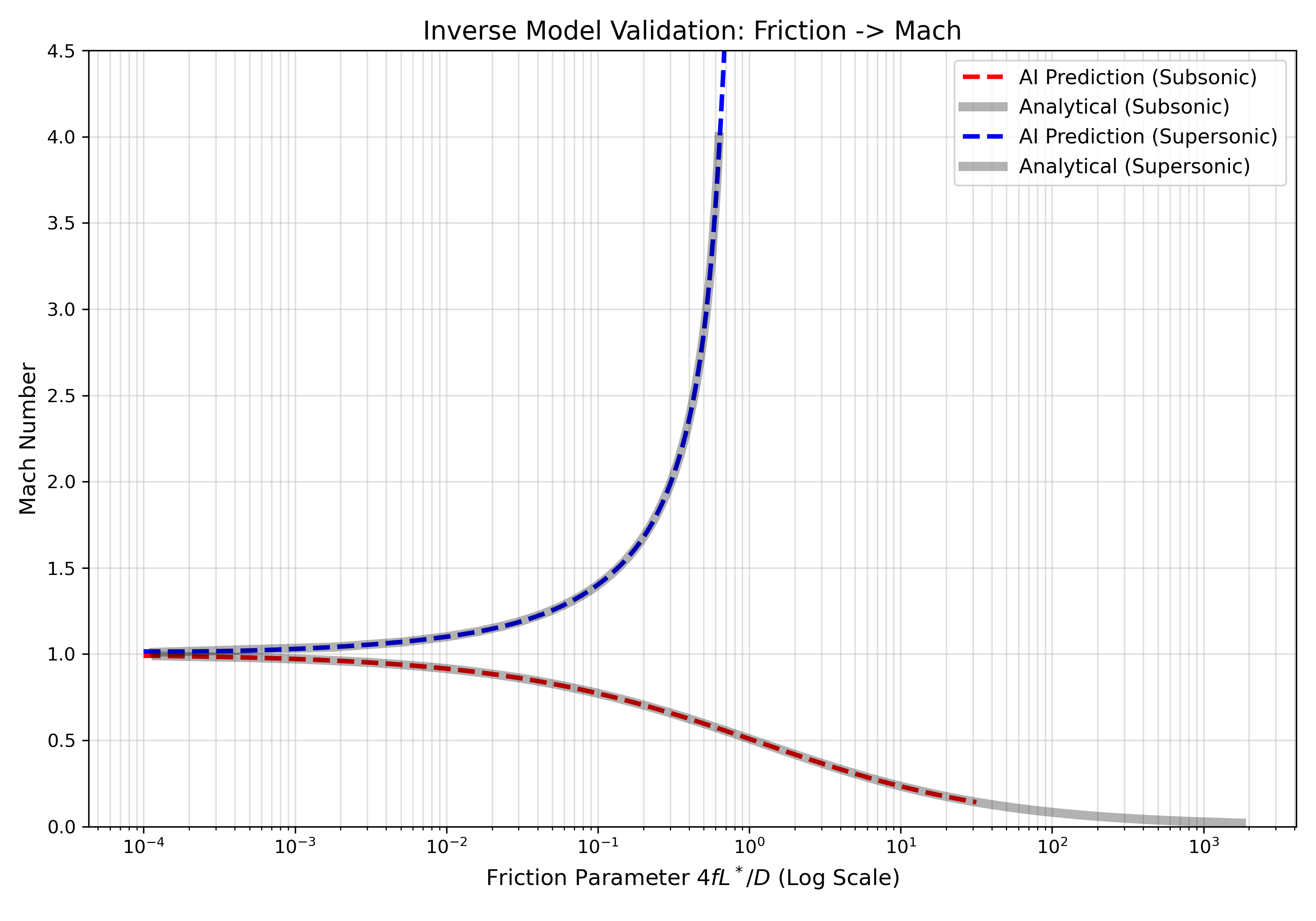}
    \caption{Comparison between the analytical solution (gray solid line) and the Deep Neural Network predictions for the subsonic (red dashed) and supersonic (blue dashed) regimes. The problem is bifurcated at the choking point ($M=1, \frac{4fL^*}{D}=0$). The supersonic branch correctly predicts the sharp increase in Mach number as the friction parameter approaches the theoretical limit (approx. 0.82 for $\gamma=1.4$), while the subsonic branch accurately captures the asymptotic behavior at low Mach numbers. The use of a logarithmic scale for the abscissa highlights the model's robustness across varied flow conditions.}
    \label{fig:inverse_fanno}
\end{figure}

\subsubsection{Thermodynamic Consistency and Case Study}
\begin{figure}[H]
    \centering
    \includegraphics[width=0.8\textwidth]{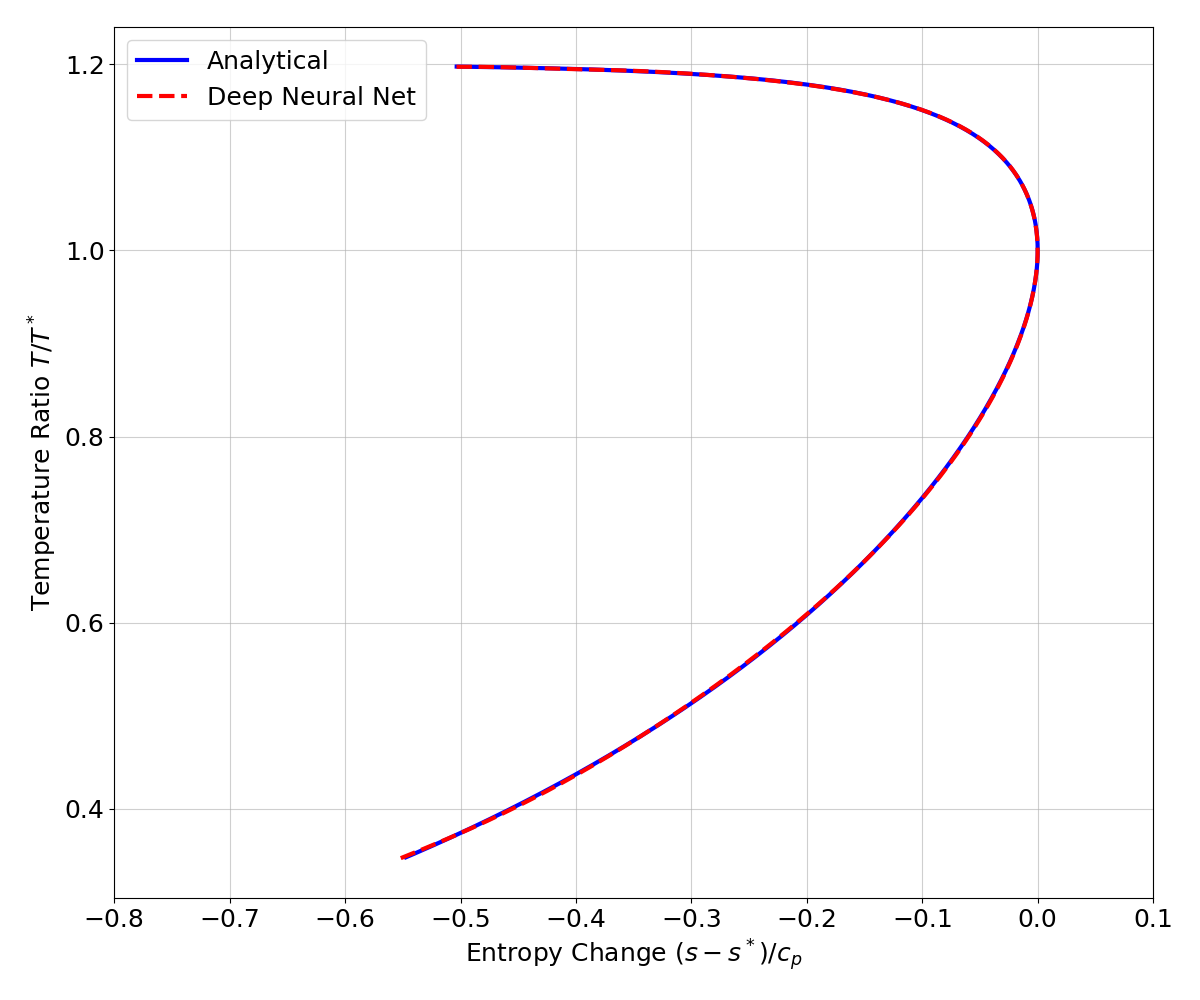}
    \caption{The Fanno line on a T-s diagram generated by the neural network. The curve correctly shows the entropy maximization point at the sonic state.}
    \label{fig:fanno_ts}
\end{figure}

\begin{figure}[H]
    \centering
    \includegraphics[width=0.8\textwidth]{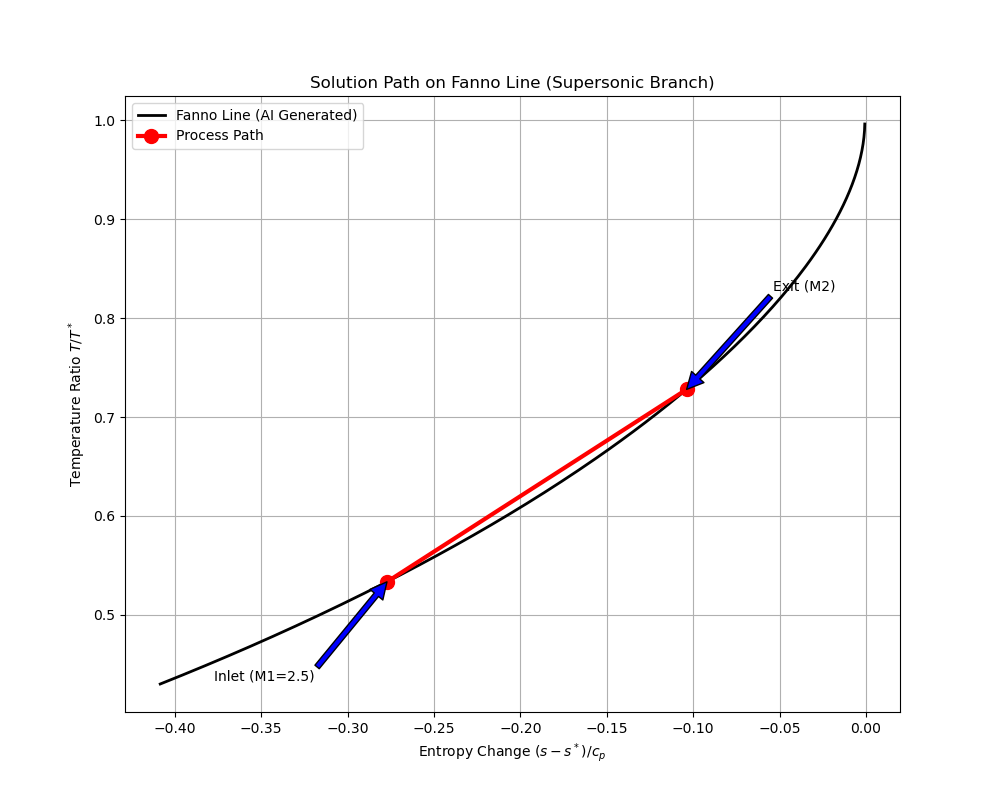}
    \caption{Visualization of a process path. A supersonic flow at M=2.5 proceeds along the Fanno line towards the exit state, computed entirely by the AI surrogate.}
    \label{fig:fanno_solution}
\end{figure}

Finally, we verify the thermodynamic consistency of the model by plotting the Fanno line on a Temperature-Entropy ($T-s$) diagram (Figure \ref{fig:fanno_ts}). The entropy change is calculated using the AI-predicted $P$ and $T$:
\begin{equation}
    \frac{s-s^*}{c_p} = \ln(T/T^*) - \frac{\gamma-1}{\gamma} \ln(P/P^*)
\end{equation}
The perfect alignment of the Fanno line confirms that the model respects the Second Law of Thermodynamics. Furthermore, Figure \ref{fig:fanno_solution} visualizes a specific case study: a supersonic flow at $M=2.5$ entering a duct. The AI solver computes the path along the Fanno line towards the sonic state, providing a clear visual aid for students.

\subsection{Oblique Shocks: The \texorpdfstring{$\beta-\theta-M$}{beta-theta-M} Manifold}

The interaction of a supersonic flow with a concave corner generates an oblique shock wave, a fundamental phenomenon in compressible aerodynamics. This interaction is governed by the strictly non-linear $\beta-\theta-M$ relation, which couples the upstream Mach number ($M_1$), the shock wave angle ($\beta$), and the flow deflection angle ($\theta$).

For a calorically perfect gas with a specific heat ratio $\gamma$, the governing equation is given by:

\begin{equation}
    \tan \theta = 2 \cot \beta \left[ \frac{M_1^2 \sin^2 \beta - 1}{M_1^2 (\gamma + \cos 2\beta) + 2} \right]
    \label{eq:beta_theta_m}
\end{equation}

Equation \ref{eq:beta_theta_m} defines a complex, non-bijective surface (manifold) in 3D space. While solving for the deflection angle $\theta$ given $M_1$ and $\beta$ is a direct algebraic calculation, the inverse problem—finding the shock angle $\beta$ for a specific geometry ($\theta$) and flight condition ($M_1$)—is implicit and requires iterative numerical methods.

The solution space is constrained by critical physical boundaries that standard regression models frequently fail to capture without physics-informed constraints:

\begin{itemize}
    \item The Mach Wave Limit ($\beta \to \mu$): As the deflection angle approaches zero ($\theta \to 0$), the shock wave weakens to an isentropic Mach wave. The shock angle converges to the Mach angle $\mu = \arcsin(1/M_1)$. Pure data-driven models often struggle here, predicting non-zero shock strengths where no discontinuity should exist.
    
    \item The Detachment Limit ($\theta_{max}$): For any given Mach number, there exists a maximum deflection angle $\theta_{max}$ beyond which Equation \ref{eq:beta_theta_m} yields no real roots. Physically, this corresponds to shock detachment, where the oblique shock transforms into a curved bow shock. 
    
    \item Solution Duality (Strong vs. Weak): For any valid pair of $M_1$ and $\theta < \theta_{max}$, there are two mathematical solutions for $\beta$: the \textit{weak solution} (prevalent in external aerodynamics) and the \textit{strong solution} (typically found in internal flows or high-backpressure scenarios).
\end{itemize}

Modeling this manifold presents a unique challenge for neural networks. A standard Mean Squared Error (MSE) loss function tends to average the weak and strong solutions near the $\theta_{max}$ turning point, resulting in physically invalid predictions that lie off the manifold. Furthermore, the steep gradients near the detachment point require high sampling density to prevent "corner-cutting" by the regression surface.

\subsubsection{ML Strategy: Physics-Informed Anchors}

Standard data-driven regression models often struggle to enforce exact boundary conditions, leading to non-zero residuals at theoretical limits. In the context of oblique shocks, a naive neural network might predict a small, non-zero deflection angle ($\theta \neq 0$) even when the shock angle corresponds to a simple Mach wave. To eliminate these unphysical artifacts, we employed a "Physics-Informed Anchor" strategy.

The training dataset was systematically augmented with synthetic boundary points derived from the limiting analytical condition where the shock strength vanishes. This limit is defined by the Mach angle $\mu$:

\begin{equation}
    \beta_{min} = \mu = \arcsin\left(\frac{1}{M}\right)
\end{equation}

At this specific angle, the flow undergoes no deflection and no compression. We injected a dense set of "anchor" pairs $(M, \beta = \mu, \theta = 0)$ into the training corpus. These anchors serve as hard constraints within the loss landscape, forcing the neural network to "pin" the regression surface to the correct physical origin ($ \theta = 0 $).

By explicitly including these theoretical zeros, the model learns to asymptotically approach the Mach wave limit rather than crossing it or floating above it. This ensures that for any valid Mach number $M$, the predicted deflection $\theta$ converges exactly to zero as $\beta \to \mu$, thereby preserving the thermodynamic consistency of the solution space.

\subsubsection{Results: Reconstruction of the Shock Manifold}

\begin{figure}[H]
    \centering
    \includegraphics[width=1.0\textwidth]{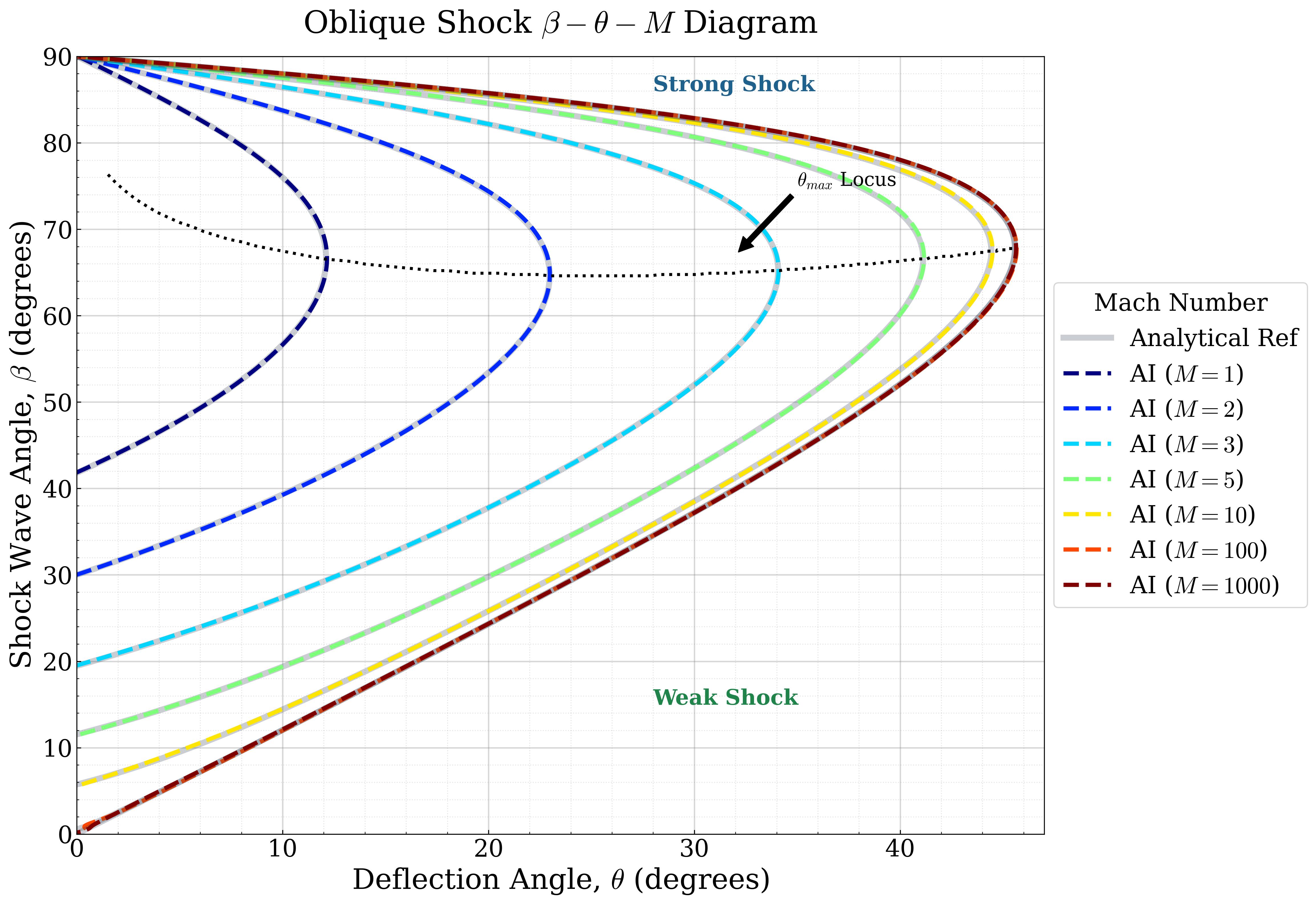}
    \caption{The AI-generated $\beta-\theta-M$ diagram. The model (dashed lines) accurately overlays the analytical theory (solid lines), respecting the physical boundary conditions at the Mach wave limits.}
    \label{fig:beta_theta}
\end{figure}

Figure \ref{fig:beta_theta} presents the superposition of the AI-predicted shock polars against the analytical exact solution. The neural network demonstrates high-fidelity agreement across the entire supersonic regime, effectively reconstructing the complex topology of the $\beta-\theta-M$ manifold.

Three key performance indicators are evident in these results:

\begin{enumerate}
    \item Resolution of Solution Duality: The model successfully captures the bifurcation of the solution space. It distinguishes between the \textit{weak solution branch} (lower $\beta$), which is stable and predominant in most external aerodynamic applications, and the \textit{strong solution branch} (higher $\beta$), which corresponds to downstream high-pressure conditions. The transition between these regimes is smooth and continuous, indicating that the network has learned the underlying trigonometric coupling rather than simply memorizing points.
    
    \item Locus of Maximum Deflection ($\theta_{max}$): The peaks of the curves, representing the critical detachment angle $\theta_{max}$ for each Mach number, are identified with high precision. This is computationally significant, as the gradient of the function $\partial \theta / \partial \beta$ becomes zero at these points, historically causing instability in standard regression techniques. The AI model accurately delimits the boundary between attached oblique shocks and detached bow shocks.
    
    \item Boundary Consistency: As a result of the physics-informed anchoring strategy, the predicted curves converge exactly to the theoretical Mach wave limits ($\beta \to \mu$) as the deflection angle approaches zero. The curves exhibit no unphysical "floating" or residual errors at the x-axis, preserving thermodynamic consistency at the onset of the shock.
\end{enumerate}

Overall, the AI solver replicates the analytical behavior with a mean absolute error (MAE) of less than 1\%, validating its potential as a surrogate model for rapid shock calculations.

\subsection{Normal Shocks in Convergent-Divergent Nozzles}

The operation of a Convergent-Divergent (C-D) nozzle is governed by the pressure ratio across it. When the back pressure ($P_b$) is low enough to choke the throat ($M_t=1$) but high enough to prevent fully supersonic flow at the exit, a normal shock wave forms inside the divergent section. Locating this shock is a classic and computationally demanding inverse problem in compressible flow education.

The flow physics involves a sequential chain of isentropic expansion, non-isentropic shock compression, and isentropic subsonic diffusion. The fundamental difficulty lies in relating the known boundary conditions (Area Ratio $A_e/A_t$ and Pressure Ratio $P_b/P_{01}$) to the unknown shock location ($A_s/A_t$).

The governing relationship links the total pressure loss across the shock to the geometric area constraints. For a perfect gas, the ratio of stagnation pressures across a normal shock is inversely proportional to the ratio of critical throat areas:

\begin{equation}
    \frac{P_{02}}{P_{01}} = \frac{A_t}{A^*_{2}}
    \label{eq:total_pressure_loss}
\end{equation}

Where $P_{01}$ is the inlet total pressure, $P_{02}$ is the exit total pressure (post-shock), $A_t$ is the physical throat area, and $A^*_{2}$ is the virtual throat area required for the flow to re-choke after the shock.

The exit static pressure $P_e$ (which equals $P_b$ for subsonic exit flow) is related to the shock location via the exit Mach number $M_e$:

\begin{equation}
    \frac{P_b}{P_{01}} = \underbrace{\left( \frac{P_e}{P_{02}} \right)}_{\text{Isentropic Subsonic Exit}} \times \underbrace{\left( \frac{P_{02}}{P_{01}} \right)}_{\text{Shock Strength}}
    \label{eq:chain_rule}
\end{equation}

Since the shock strength depends entirely on the Mach number immediately upstream of the shock ($M_{s1}$), and $M_{s1}$ is determined by the area ratio $A_s/A_t$, Equation \ref{eq:chain_rule} becomes a transcendental function of the shock area $A_s$. Solving for $A_s$ typically requires an iterative "shooting method": guessing a shock location, calculating the resulting pressure loss, and refining the guess until the calculated exit pressure matches the given back pressure.

\subsubsection{ML Strategy: Design Space Mapping}

To bypass the iterative computational cost and provide immediate design feedback, we formulated the shock location problem as a direct regression task. We trained a Deep Neural Network (DNN) to approximate the inverse function $\mathcal{F}^{-1}$:

\begin{equation}
    \frac{A_s}{A_t} \approx \text{DNN}\left( \frac{A_e}{A_t}, \frac{P_b}{P_{01}} \right)
\end{equation}

The network maps the design constraints (Geometry and Back Pressure) directly to the physical response (Shock Area). This approach effectively "flattens" the iterative physics solver into a single forward-pass operation, reducing calculation time from milliseconds (iterative) to microseconds (neural inference).

\subsubsection{Visualization: The Shock Location Heatmap}

\begin{figure}[H]
    \centering
    \includegraphics[width=0.9\textwidth]{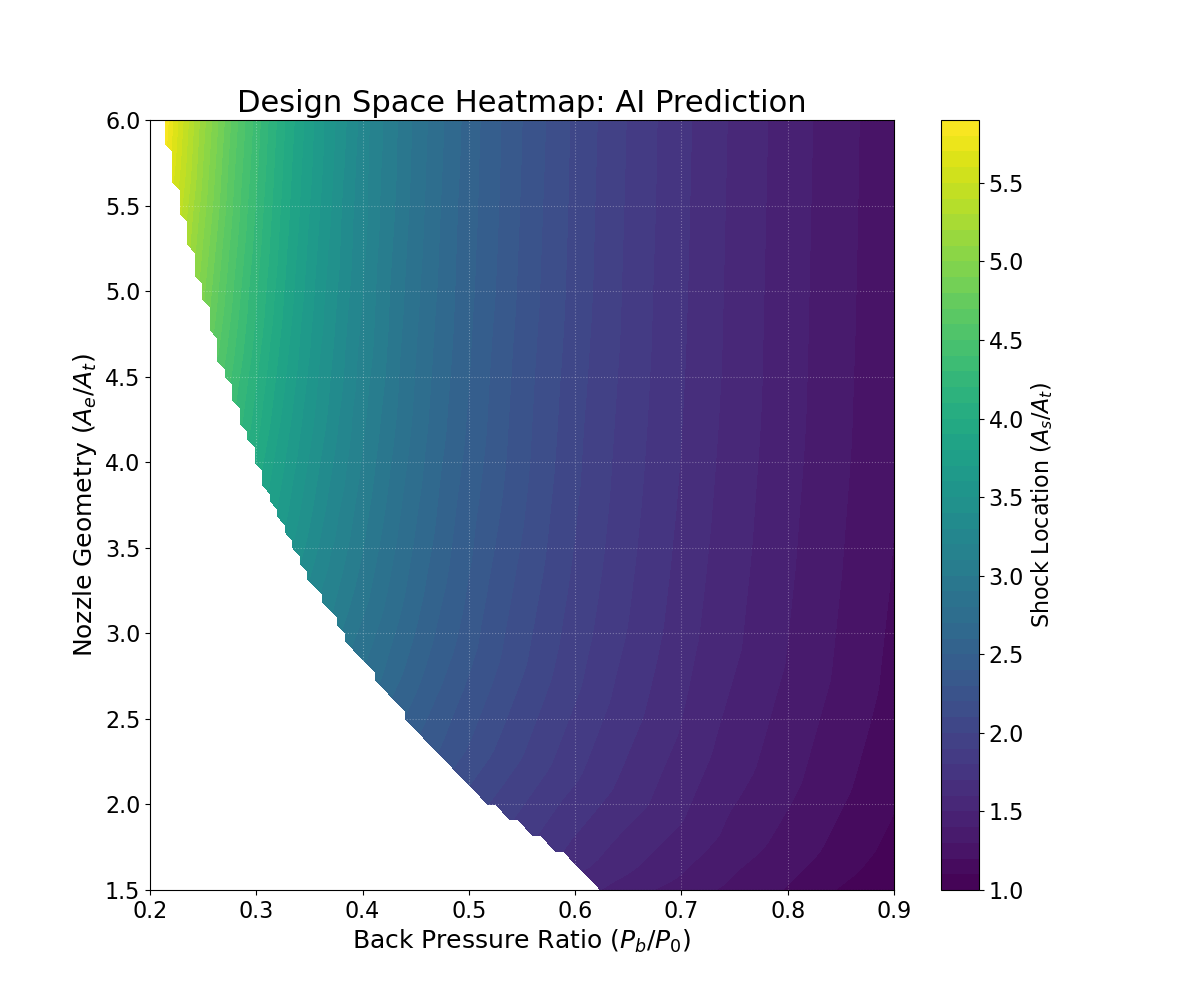}
    \caption{Design Space Heatmap generated by the AI model. It maps the relationship between nozzle geometry (y-axis), back pressure (x-axis), and the resulting shock location (color gradient).}
    \label{fig:nozzle_heatmap}
\end{figure}

The speed of the AI solver enables the real-time generation of a "Design Heatmap" (Figure \ref{fig:nozzle_heatmap}), which visualizes the entire solution space of the nozzle. This diagram plots the Nozzle Area Ratio ($A_e/A_t$) on the x-axis and the Back Pressure Ratio ($P_b/P_{01}$) on the y-axis, with the color contour representing the shock location ($A_s/A_t$).

This visualization highlights several critical flow regimes:
\begin{itemize}
    \item The Valid Shock Domain: The colored region represents the specific combinations of geometry and pressure where a normal shock can physically exist inside the nozzle.
    \item he Throat Limit ($A_s \to A_t$): As back pressure increases (moving up the y-axis), the shock is pushed upstream towards the throat. The heatmap clearly shows the boundary where the shock vanishes and the nozzle unchokes (First Critical Point).
    \item The Exit Limit ($A_s \to A_e$): As back pressure decreases, the shock moves downstream. The lower boundary of the heatmap corresponds to the "Third Critical Point," where the normal shock sits exactly at the exit plane. Below this line, the flow becomes over-expanded with oblique shocks outside the nozzle.
\end{itemize}

By exploring this heatmap, students can intuitively grasp how sensitive the shock position is to small changes in back pressure, a concept that is often obscured by single-point algebraic calculations.

\subsubsection{Results: Pressure Distribution Reconstruction}

\begin{figure}[H]
    \centering
    \includegraphics[width=1.0\textwidth]{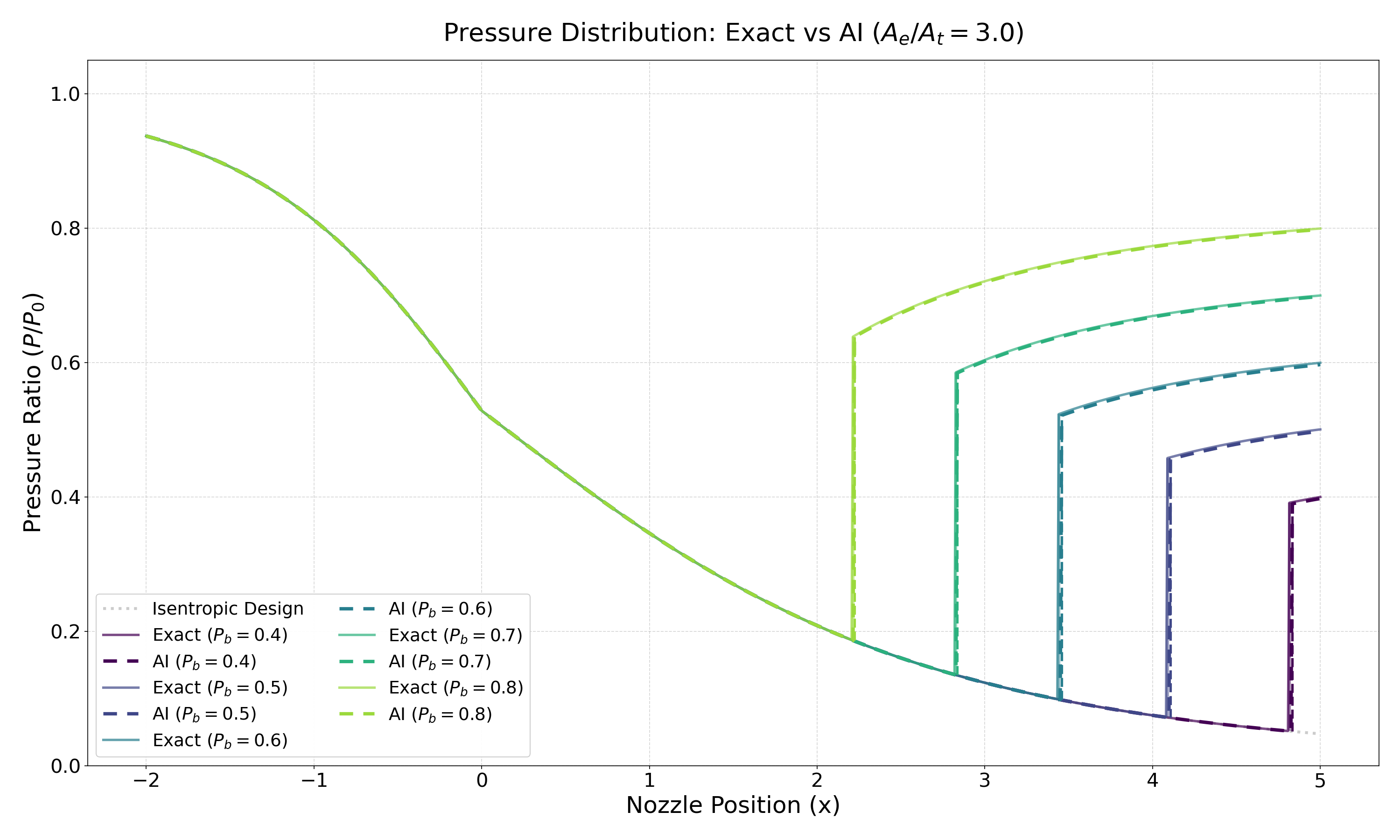}
    \caption{Pressure distribution along the nozzle axis. The solid lines represent the exact analytical solution, while dashed lines represent the AI reconstruction. The model accurately predicts the shock position for various back-pressures.}
    \label{fig:nozzle_pressure}
\end{figure}

We coupled the AI shock locator with an analytical reconstructor to generate the pressure distribution along the nozzle (Figure~\ref{fig:nozzle_pressure}). This figure compares the exact analytical solution (solid lines) with the AI-based reconstruction (dashed lines) for the static pressure ratio, $P/P_{0}$, along the nozzle axis. The results correspond to a fixed nozzle geometry with an area ratio of $A_{e}/A_{t} = 3.0$, evaluated under various back-pressure conditions ranging from $P_{b}/P_{0} = 0.4$ to $0.8$.

The plot clearly illustrates the sequence of flow processes: isentropic expansion in the convergent and initial divergent sections, followed by a sharp, discontinuous pressure rise across a normal shock wave, and finally, isentropic subsonic compression to match the exit boundary condition. As expected physically, increasing the back pressure forces the shock wave to move upstream towards the nozzle throat. The AI model accurately predicts this shift in shock position for all tested back pressures. The precise capture of the shock's location and strength leads to a near-perfect reconstruction of the entire pressure profile, validating the efficacy of the proposed machine learning approach for this complex inverse problem.

\subsection{Unsteady Shock Tubes: The Riemann Problem}

The shock tube problem constitutes a fundamental benchmark in compressible fluid dynamics, representing the simplest case of unsteady, one-dimensional wave interaction. Physically, it consists of a high-pressure "driver" gas (Region 4) separated from a low-pressure "driven" gas (Region 1) by a diaphragm. Upon the instantaneous rupture of the diaphragm, a shock wave propagates into the low-pressure region, while an expansion fan propagates back into the high-pressure region.

This configuration is mathematically governed by the 1D Euler equations. The central challenge lies in determining the strength of the generated shock wave, defined by the pressure ratio $P_{2}/P_{1}$, given the initial diaphragm pressure ratio $P_{4}/P_{1}$. These two ratios are linked via the fundamental shock tube equation, a highly non-linear implicit function derived by matching pressure and velocity across the contact surface:

\begin{equation}
    \frac{P_4}{P_1} = \frac{P_2}{P_1} \left[ 1 - \frac{(\gamma_4-1)(a_1/a_4)(P_2/P_1 - 1)}{\sqrt{2\gamma_1 \left( 2\gamma_1 + (\gamma_1+1)(P_2/P_1 - 1) \right)}} \right]^{-\frac{2\gamma_4}{\gamma_4-1}}
    \label{eq:shock_tube_implicit}
\end{equation}

Where $a$ denotes the speed of sound and $\gamma$ is the specific heat ratio. Equation \ref{eq:shock_tube_implicit} cannot be solved analytically for $P_2/P_1$. Traditional Computational Fluid Dynamics (CFD) approaches solve this via iterative root-finding algorithms (e.g., Newton-Raphson) at every time step, or by computationally expensive finite-volume time-marching schemes.

\subsubsection{ML Strategy: The Hybrid Solver}

To address the implicit nature of Equation \ref{eq:shock_tube_implicit} without incurring the computational cost of iterative solvers, we developed a "Hybrid AI-Analytical Solver." 

In this architecture, a lightweight Deep Neural Network serves as a direct function approximator for the inverse of the shock tube equation. The network takes the initial conditions ($P_4/P_1, T_4/T_1$) as inputs and instantly predicts the shock strength $P_2/P_1$.

Crucially, the AI is not tasked with predicting the entire flow field pixel-by-pixel, which often leads to blurred discontinuities in pure computer vision approaches. Instead, the AI output ($P_2/P_1$) is fed into a deterministic analytical reconstruction module. Once $P_2$ is known, the properties in all other regions (Region 2 behind the shock and Region 3 behind the contact surface) become explicit algebraic calculations:

\begin{equation}
    T_2 = T_1 \left[ \frac{P_2}{P_1} \left( \frac{\frac{\gamma+1}{\gamma-1} + \frac{P_2}{P_1}}{1 + \frac{\gamma+1}{\gamma-1} \frac{P_2}{P_1}} \right) \right]
    \label{eq:temp_ratio}
\end{equation}

This hybrid approach combines the inference speed of neural networks (microseconds) with the infinite precision of analytical gas dynamics, ensuring exact conservation of mass, momentum, and energy.

\subsubsection{Results: Flow Field and Wave Propagation}

The capabilities of the hybrid solver are demonstrated through the reconstruction of instantaneous property profiles and space-time wave histories.

\begin{figure}[H]
    \centering
    \includegraphics[width=1.0\textwidth]{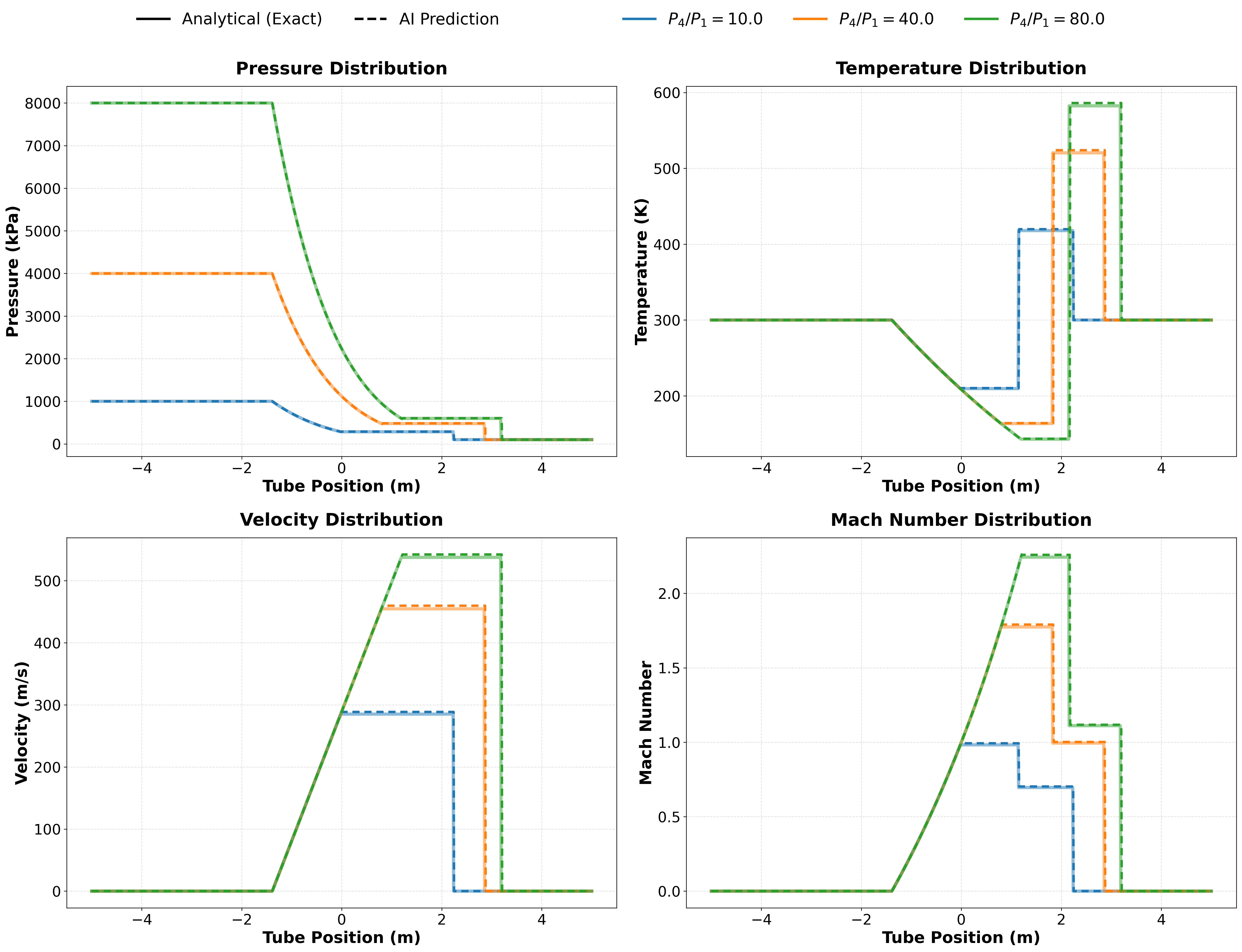}
    \caption{Shock tube flow field reconstruction at a specific time step ($t=4ms$). Comparison of AI prediction (dashed) vs. Analytical solution (solid) for pressure ratios of 10, 40, and 80. Note the accurate capture of the contact surface discontinuity in the temperature plot.}
    \label{fig:shock_tube_profile}
\end{figure}

Figure \ref{fig:shock_tube_profile} compares the instantaneous spatial distribution of Pressure, Temperature, Velocity, and Mach number for three distinct driver pressure ratios. The solver exhibits near-perfect agreement with the exact solution.

A critical test of any shock tube solver is the resolution of the \textit{Contact Surface} (the interface between the driver gas and the driven gas). Across this interface:
\begin{itemize}
    \item Pressure and Velocity must be continuous ($P_2 = P_3$, $u_2 = u_3$).
    \item Temperature and Density are discontinuous ($T_2 \neq T_3$, $\rho_2 \neq \rho_3$).
\end{itemize}
As highlighted in the temperature plot of Figure \ref{fig:shock_tube_profile}, the AI-driven model successfully captures this subtle physical phenomenon. While the pressure trace is flat across the middle of the tube, the temperature trace shows a sharp step change, confirming that the model effectively distinguishes between the shock wave (which compresses and heats) and the contact surface (which separates gases of different entropies).

\begin{figure}[H]
    \centering
    \includegraphics[width=1.0\textwidth]{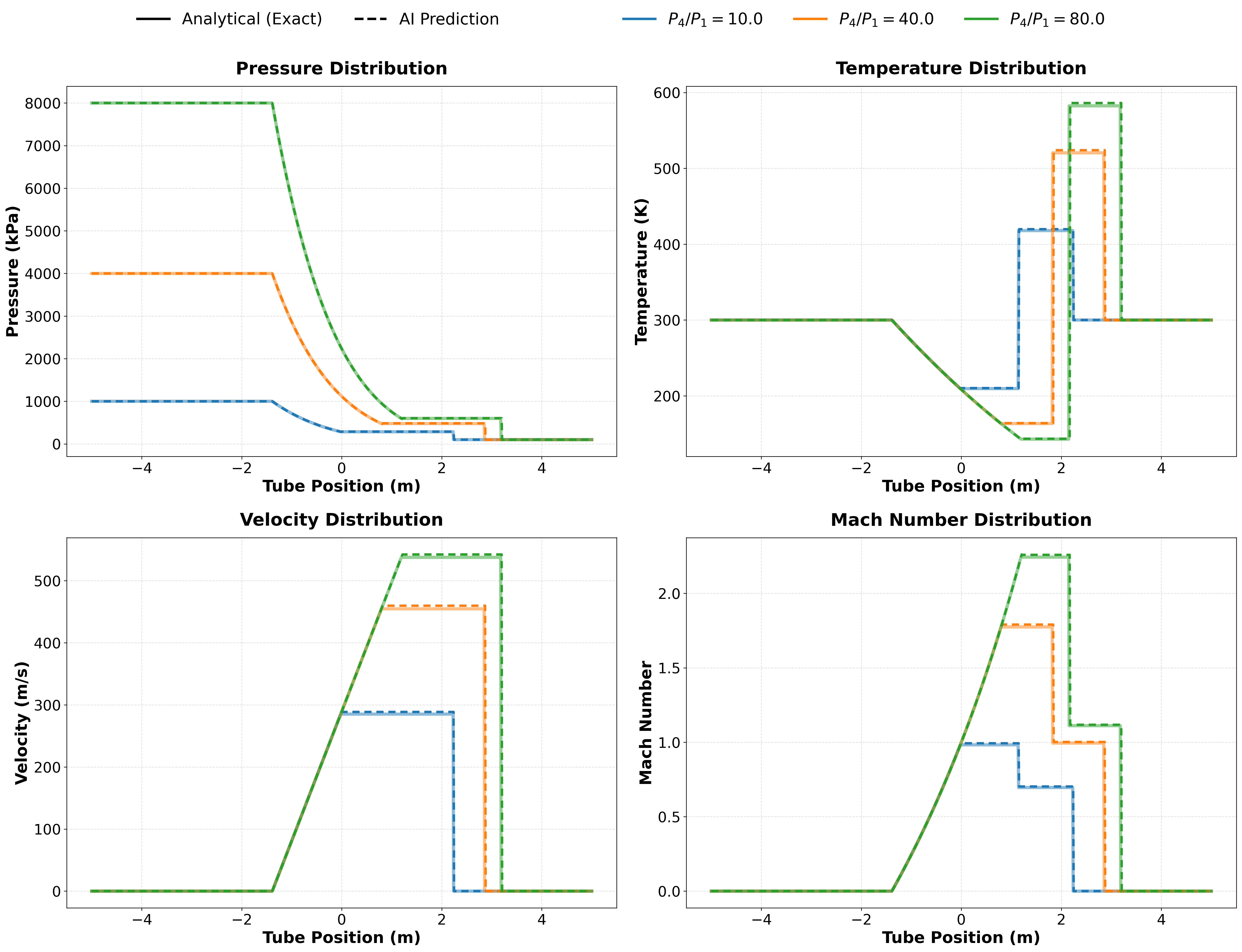}
    \caption{Space-time ($x-t$) contour plots generated by the AI model. Left: Pressure field showing the shock and expansion fan. Right: Temperature field clearly highlighting the path of the contact surface (the sharp interface between cold and hot gas).}
    \label{fig:shock_tube_xt}
\end{figure}

Figure \ref{fig:shock_tube_xt} visualizes the complete temporal evolution of the flow field. Standard CFD requires time-marching integration to generate such a plot. In contrast, our AI surrogate generates this entire solution space in milliseconds by vectorizing the analytical reconstruction.

The contour plot clearly delineates the three primary wave features:
\begin{enumerate}
    \item The Shock Wave: Propagating to the right into the quiescent medium (Region 1).
    \item The Contact Surface: Following the shock at a lower velocity, marking the boundary between the hot compressed gas and the cold expanded gas.
    \item The Expansion Fan: Propagating to the left into the high-pressure driver (Region 4), characterized by a smooth, continuous gradient of properties rather than a sharp discontinuity.
\end{enumerate}

The sharp definition of the expansion fan "head" (moving at $-a_4$) and "tail" (moving at $u_3 - a_3$) demonstrates the solver's ability to handle mixed supersonic/subsonic flow regimes simultaneously.

\section{Conclusion}

This study presents a paradigm shift in the computational treatment of fundamental compressible flow problems. We have demonstrated that Deep Learning (DL) can function as far more than a black-box tool for big data; it is a precision instrument for solving the non-linear, implicit inverse problems that characterize gas dynamics. By developing a suite of physics-aware neural networks, we have effectively bridged the gap between static textbook equations and dynamic, real-time simulation.

\subsection{Summary of Technical Contributions}
The core innovation of this work lies in the tailored application of machine learning architectures to specific physical regimes. Rather than applying a generic regression model, we introduced distinct strategies for each flow class:
\begin{itemize}
    \item Domain Decomposition: For Fanno and Rayleigh flows, splitting the solution space into distinct subsonic and supersonic "expert" networks resolved the issue of non-bijective functions, enabling instant friction and heat transfer predictions without iterative root-finding.
    \item Physics-Informed Anchors: In the modeling of Oblique Shocks, the injection of theoretical boundary points (Mach waves) into the training corpus successfully constrained the solution manifold, ensuring thermodynamic consistency near the limit of zero deflection.
    \item Hybrid Architectures: For the Unsteady Shock Tube and Convergent-Divergent Nozzle problems, we demonstrated the power of a hybrid solver. By using the neural network solely to resolve the implicit algebraic step (e.g., finding the shock strength $P_2/P_1$) and relying on analytical gas dynamics for field reconstruction, we achieved the speed of AI while maintaining exact conservation of mass, momentum, and energy.
\end{itemize}

\subsection{Pedagogical and Engineering Implications}
The transition from traditional look-up tables and iterative solvers to AI surrogates offers profound benefits for both education and industry.

From a pedagogical perspective, this approach modernizes the curriculum. Students are often bogged down by the arithmetic mechanics of "shooting methods" or interpolating values from Gas Tables. The "Design Heatmaps" and real-time $x-t$ diagrams generated by our models allow learners to bypass these arithmetic hurdles and focus on the underlying physics. They can interactively visualize how a normal shock moves through a nozzle in response to back-pressure changes or how a contact surface propagates in a shock tube, fostering a deeper, intuitive understanding of the flow field.

In an engineering context, these surrogates represent a powerful tool for preliminary design and optimization. The ability to solve inverse problems in microseconds—orders of magnitude faster than conventional CFD—allows for the exploration of vast design spaces. An engineer can now instantly assess millions of duct configurations or flight conditions to identify optimal operational windows before committing to expensive high-fidelity simulations.

\subsection{Future Outlook}
While this work focuses on 1D inviscid flows, the methodology establishes a robust framework for more complex applications. Future work will extend these physics-informed strategies to 2D viscous interactions and reacting flows. Ultimately, the integration of such lightweight, high-precision AI models into standard CFD solvers could serve as efficient preconditioners or sub-grid scale models, further accelerating the simulation of complex aerodynamic systems.

\subsection*{Source Codes}
The inference scripts are available at https://github.com/Ehsan-Roohi/Gas-Dynamics-.

\end{document}